# Theoretical Efficiency Comparison between Carrier Multiplication and Down-Conversion 3rd Generation Solar Cell Designs

Ze'ev R. Abrams, Avi Niv, Chris Gladden, Majid Gharghi and Xiang Zhang

*NSF Nanoscale Science and Engineering Center (NSEC), 3112 Etcheverry Hall, University of California, Berkeley, California 94720, USA*

*Materials Science Division, Lawrence Berkeley National Laboratory, 1 Cyclotron Road, Berkeley, California 94720, USA*

**Abstract**

Methods of exceeding the detailed balance limit for a single junction solar cell have included down-converting high energy photons to produce two photons; and carrier multiplication, whereby high energy photons produce more than one electron-hole pair. Both of the methods obey the conservation of energy in similar ways, and effectively produce a higher current in the solar cell. Due to this similarity, it has been assumed that there is no thermodynamic difference between the two methods. Here, we compare the two methods using a generalized approach based on Kirchhoff's law of radiation and develop a new model for carrier multiplication. We demonstrate that there is an entropic penalty to be paid for attempting to accomplish all-in-one splitting in carrier multiplication systems, giving a small thermodynamic - and therefore *efficiency* - advantage to spectral splitting prior to reaching the solar cell. We show this analytically using a derivation of basic thermodynamic identities; numerically by solving for the maximal efficiency; and generally using heat-generation arguments. Our result modifies the existing literature on entropy generation limits in solar cells, and creates a new distinction among 3rd generation photovoltaic technologies.

**1. Introduction**

The 31% Detailed Balance (DB) limit for a single junction solar cell by Shockley and Queisser [1] is based upon the reduction of the Ultimate Efficiency (UE) limit of 44% due to entropic losses [2]. The UE limit is based on the UE hypothesis [1] of a single electron-hole (e-h) pair per incoming photon. However it was understood from the outset [1,3] that higher energy photons may have a higher-than-unity quantum efficiency of producing e-h pairs, resulting in carrier multiplication that would result in efficiencies higher than those determined by the UE hypothesis. Carrier multiplication (CM), either Multiple Exciton Generation (MEG) or impact ionization [4,5,6,7], has been amongst the most prominent "3rd Generation" techniques considered for exceeding the current efficiency limits for a single junction cell [8]. Few CM materials have been found that can produce a perceptible efficiency increase [6,9,10], with recent works questioning the plausibility of realistic efficiency increases using CM [9,11].

The CM method relies on the increase in short circuit current of the solar cell, $I_{sc}$, which increases proportionately to the ratio of photons with energies above twice the band-gap. In an analogous concept, a Down-Converting (DC) system placed adjacent to a solar cell [12] can produce a similar increase in current by quantum-splitting a higher energy photon into two (or more) smaller energy photons, thus ensuring the principal of energy conservation. We have recently analyzed the DC process in depth [13], including a complete thermodynamic description of the process, and have demonstrated a slight entropic *gain* to the efficiency of the underlying solar cell due to the spectral splitting of the DC layer.

Since the two methods are so similar, it has been assumed that there is no difference in efficiency limits between them [8], with the increase in efficiency being entirely dependent upon the $I_{sc}$ increase. However, we have shown that the DC process also benefits from an entropic gain term that increases the open-circuit voltage, $V_{oc}$, when analyzed at radiation-flux equilibrium [13]. This increase can be found as a function of the increase in free energy, as defined by the Gibbs free energy [8,16] of a system with energy $U$, temperature $T$, and entropy $S$:

$$\mu_{oc} = Gibbs = U - T \times S \tag{1}$$

with $\mu_{oc}$ being the chemical potential at steady-state, open-circuit conditions being proportionate to $V_{oc}$ ($\mu_{oc}=qV_{oc}$, with $q$ being the elementary charge). For a DC system, we showed that the lower entropy production in the solar cell results in a slightly higher total efficiency, since the efficiency is proportionate to the product $I_{sc} \times V_{oc}$. An obvious question to address then is whether there is any thermodynamic difference between the DC and CM schemes that will manifest itself in the free energy of the system? To compare the systems, we develop a first-order analysis for CM based on a generalization of Kirchhoff's law of radiation that is different from the currently accepted CM/MEG models [8,17,18].

Our manuscript first presents this analysis method for regular, DC and CM solar cells, followed by the open- and short-circuit conditions for each system. This method allows us to directly isolate the chemical potential using the Ruppel and Würfel photon flux method [19], which is a rephrasing of Kirchhoff's law of radiation [16] using the photon *rate* equations (instead of power) based on the van-Roosbroeck-Shockley (vRS) relation [1,20]. We show the entropic difference between the two methods as a function of the effective addition of number of photons. In section 3 we solve these equations numerically, showing the maximal efficiency for each method, when compared to a regular solar cell. In section 4 we analyze the systems from a heat-generation perspective, with the thermalization of electrons generating heat losses within the solar cell. With each of these methods, a clear advantage is seen for the DC method, as a representative of the spectral splitting methodology, as compared with the CM technique, which attempts to package the spectral splitting capabilities within the solar cell itself. Finally, in section 5 we discuss the consequences of these results in terms of general thermodynamics [21,22], comparing our results with other models of CM/MEG [17,18,23,24], as well as its implications for other spectral splitting methods.

## 2. Analytical Formulas for Solar Cells

### 2.1. *Regular Solar Cell*

To first order, the efficiency of a solar cell with a single band-gap can entirely be contained within the relation between incoming blackbody photons from the sun and their re-emission as blackbody photons from the solar cell itself [8,16]. The photon flux emitted from the sun, a blackbody at temperature $T_S$=6000 K, and absorbed by a semiconductor with band-gap $E_g$, is given by:

$$\dot{N}_{ph}^{Sun} = g \times \varepsilon \times \int_{E_g}^{\infty} \frac{E^2 dE}{e^{E/kT_S} - 1} \qquad (2)$$

with $g$ being a constant ($g=2/h^3c^2$, in units of $1/[eV^3][sec]$), and with a geometrical factor inserted into the generalized étendue $\varepsilon=\Omega_S=6.85\times10^{-5}$ sr, which includes the solid angle of the incident sunlight. Any additional concentration, $C$, to the system can be directly inserted into the étendue term [25], with $\varepsilon_{concentrated}=C\times\Omega_S$, assuming the concentration is maximized for a spherical solar cell illuminated by full-spherical illumination at $4\pi$ sr [1]. In contrast, the photons emitted from a semiconducting solar cell at ambient temperature $T_o=300$ K, with chemical potential $\mu$ follow a modified emission rate using the modified vSR relation. At open-circuit conditions this is:

$$\dot{N}_{ph}^{cell} = g \times 4\pi \times \int_{E_g}^{\infty} \frac{E^2 dE}{e^{(E-\mu_{oc})/kT_o} - 1} \qquad (3)$$

The output emission is typically multiplied by the square of the refractive index of the solar cell material ($n^2$), which is here taken as unity, and is assumed to radiate isotropically ($\varepsilon=4\pi$). The current of the solar cell can be simplified as being directly related to the photon fluxes, with $I = q\dot{N}_{ph}^{Sun} - q\dot{N}_{ph}^{cell}$ [26]. Since the chemical potential is directly related to the voltage extracted from the solar cell assuming such equilibrium conditions, these equations directly provide all the important equations relevant to solar cell efficiency calculations [8,16].

The chemical potential can be viewed as a "valve" whose extreme values are at short-circuit, when $\mu=0$ providing maximal current, and at open-circuit, with $\mu=\mu_{oc}$ providing zero current. Due to the temperature differences, the outgoing emission at short-circuit is 5 orders of magnitude smaller than the input, and essentially: $I \cong q\dot{N}_{ph}^{Sun}$. In contrast, at open-circuit conditions, using the Ruppel and Würfel relation [19], the outgoing emission equals the incoming absorption (Kirchhoff's law of radiation). Since $\mu_{oc}$ is an unknown that provides us with thermodynamic information [as in Eq. (1)], we can approximate $\mu_{oc}$ for $E>>kT_S$ and $E-\mu>>kT_o$, resulting in well known analytical description of $V_{oc}$ [19,27]:

$$qV_{oc}^{reg} = E_g \eta_C - kT_o \ln\left[\left(\frac{\Omega_S}{4\pi}\right)\left(\frac{T_S}{T_o}\right)\alpha_1\right] \qquad (4)$$

Where the Carnot efficiency is defined as $\eta_C=(1-T_o/T_S)=95\%$ [28] for these temperatures, and $\alpha_1$ is a small correction term [$\alpha_1 = 1 + 2kT_S/E_g + 2(kT_S/E_g)^2$], as defined by us [13] and others [25]. For $E_g>kT_S$, the bracketed term can be considered a constant as a function of $E_g$, such that using Eq. (1), we can recognize the entropy as the term: $S=k\times ln(\cdot)$ [27]. The entropy is comprised of a temperature difference component ($T_S/T_o$), a generalized étendue component ($\Omega_S/4\pi$), and the contribution of non-radiative recombination can be added as well [25,27,29].

## 2.2. Down-Conversion Solar Cell

The DC system must include the contribution of the external layer implementing the spectral splitting. As we argued in Ref. [13], the DC layer is an isolated system that can be analyzed separately from the underlying solar cell (as opposed to the circuit model used in [12,14]). We assume a DC layer placed *above* the solar cell, such that the incident light is modified before reaching the cell itself. As a result, the input spectrum into the solar cell is modified to include the external splitting of higher energy photons:

$$\dot{N}_{ph,in}^{DC} = g \times \varepsilon \times (1 + f_{DC} \times \theta) \times \int_{E_g}^{\infty} \frac{E^2 dE}{e^{E/kT_S} - 1} \tag{5}$$

The modified spectrum manifests itself as an additional fraction of $f_{DC} \times \theta$, where $0 \leq f_{DC} \leq 1$ is the efficiency of DC, and $\theta$ denotes the fraction of photons above $2E_g$:

$$\theta(E_g) \equiv \int_{2E_g}^{\infty} \frac{E^2 dE}{e^{E/kT_S} - 1} \bigg/ \int_{E_g}^{\infty} \frac{E^2 dE}{e^{E/kT_S} - 1} \tag{6}$$

and $\theta$ is a monotonically decreasing function of $E_g$.

The difference between the regular solar cell and the DC solar cell is therefore the incoming photon flux. Using Kirchhoff's law of radiation, the modified incoming flux in Eq. (5) must equal the outgoing flux at open-circuit conditions. However, since the solar cell itself is unchanged, the outgoing flux equation remains that of Eq. (3), as the DC layer does not affect the emission properties of the solar cell (excluding refractive index effects [14]). Using the same approximations used to derive $\mu_{oc}$ in Eq. (4), we can obtain for DC:

$$qV_{oc}^{DC} = qV_{oc}^{reg} + kT_o \ln(1 + f_{DC} \times \theta) \tag{7}$$

with $qV_{oc}^{reg}$ given by Eq. (4). This results in a $V_{oc}$ that is slightly *higher* for the DC system than for a regular solar cell. In terms of Kirchhoff's law of radiation, the chemical potential of the DC system must compensate for the slightly elevated number of incoming photons due to the DC splitting layer by increasing $\mu$ slightly, thereby increasing the emission from the solar cell at open-circuit. Entropically, we argued [13] that this added gain is due to the decrease in the entropy of the solar spectrum by removing the number of photons with $h\nu > 2E_g$, resulting in a slight increase in the accessible free energy of the solar cell. This can be seen by setting $f_{DC} = 1$, and noting that this gain term is simply the logarithm of the ratio of the new DC spectrum to the original solar spectrum absorbed by the cell.

## 2.3. Carrier Multiplication Solar Cell

The initial model for analyzing CM [30] assumed that the excess carrier generation could be modeled identically as the DC input current [Eq. (5)], but only contained a single emission channel following Eq. (3), exactly as in our DC model above. However, it was soon commented that the 2[nd] law of thermodynamics is broken at maximal concentration, with a net *negative* entropy produced using this model [21,22]. As a result, a modified model was introduced [17,18,16] in an attempt to comply with

Kirchhoff's law of radiation, such that the increase in photon influx induces a concurrent re-emission of photons with *different* chemical potentials (essentially with the chemical potential following the quantum efficiency model of the CM system [31]). For a single multiplication level, this output emission is:

$$\dot{N}_{ph,out}^{MEG}(original) = g \times 4\pi \left( \int_{E_g}^{2E_g} \frac{E^2 dE}{e^{(E-\mu)/kT_o} - 1} + \int_{2E_g}^{\infty} \frac{E^2 dE}{e^{(E-2\mu)/kT_o} - 1} \right) \tag{8}$$

These photons could be described as reverse-MEG, or similar to Auger recombination. The difference between models could only be seen at high concentrations and low band-gaps [23,24,32]. An additional attempt to develop a theory for CM using "first principals" of thermodynamics [33] also resulted in internal splitting levels with different chemical potentials, similar to hot-electron devices [34]. However, a problem with these models is that at open-circuit conditions, there should only be a single uniform $\mu_{oc}$ [32], as measured by the $V_{oc}$ of the solar cell.

To address this issue, we have devised a different model for the CM system. A strict application of Kirchhoff's law to a CM system would violate the conservation of flux, since more photons are emitted than absorbed. This is due to the nonlinearity of the CM system for energies above $2E_g$, whereas for photons with energy $E_g \leq h\nu \leq 2E_g$, the system remains linear. To account for this internal nonlinearity, we can split the system into two parallel systems: the first for photons with $E_g \leq h\nu \leq 2E_g$ follows the regular application of Kirchhoff's law of radiation [equating Eqs. (2) and (3) with the proper limits of integration]; and the second, replaces the absorbed light with Eq. (2) multiplied by the fraction of photons generating multiple carriers, given by $\theta$ in Eq. (6). To retain linearity in this sub-system, we must multiply the *emission* by the same factor, $\theta$.

Physically, this model can be understood by replacing the solar spectrum with monochromatic beams at two frequencies, and with the same intensity as the solar irradiance. For monochromatic photons with $h\nu = E_g$, the current is directly proportionate to the incoming beam, and the emission is equal to the absorption. However, for a beam with $h\nu = 2E_g$, all photons are converted into 2 e-h pairs, thus doubling the emitted photons as well. We can thus see that the *current* is directly proportionate to the outgoing emission flux, with both increasing linearly by a factor of 2. Therefore, the emission of photons in this sub-system, with $h\nu > 2E_g$, is linearly proportionate to the incoming flux, multiplied by the factor $\theta$.

Additionally, we can derive this linear increase of the emission by assuming a 3-level system for the CM process, with absorption and relaxation between the valence and conduction bands (lower two levels), as well as an additional component of absorption from the valence band to the top-MEG band. By assuming that no absorption can occur between the conduction and MEG bands, and assuming that there is no relaxation from the MEG band directly to the valence band (which would not produce an additional e-h pair), then the principal relaxation from the conduction band to the valence band must increase.

Applying this model, constraining the chemical potential of the system to a single $\mu$, and re-interpreting Kirchhoff's law of radiation in this case such that the increase in current due to the CM process by a factor of $\theta$ must involve a concomitant increase in emission of photons at open-circuit equilibrium conditions by the same factor of $\theta$, we can write the outgoing flux of the CM system as:

$$\dot{N}_{ph,out}{}^{MEG}(new) = g \times 4\pi \left( \int_{E_g}^{\infty} \frac{E^2 dE}{e^{(E-\mu_{oc})/kT_o} - 1} + f_{CM}\theta \times \int_{E_g}^{\infty} \frac{E^2 dE}{e^{(E-\mu_{oc})/kT_o} - 1} \right) \quad (9)$$

where we have also included a CM efficiency term: $0 \leq f_{CM} \leq 1$. Using the same derivation as before, by equating Eqs. (5) and (9), we can find the open-circuit voltage:

$$qV_{oc}{}^{CM} = qV_{oc}{}^{DC} - kT_o \ln(1 + f_{CM} \times \theta) = qV_{oc}{}^{reg} \quad (10)$$

where the difference between Eqs. (10) and Eq. (7) is the *negative* term (assuming $f_{DC}=f_{CM}$). By adding the restriction that $\mu$ throughout the CM solar cell is uniform (as appearing in the original Shockley-Queisser paper [1]), it can be seen that the penalty of attempting to split the photon flux *within* the solar cell is tantamount to losing the entropy *gain* exhibited in the DC system. The CM system's thermodynamics are therefore more similar to a regular solar cell than a DC one (see, however, section 4).

Alternatively, we can describe our model using Kirchhoff's law of radiation: we see that the *linear* contribution of internal carrier generation to the input current term by a factor of $f_{CM}\theta$ photons per second results in a directly proportionate *linear* increase in blackbody emission from the solar cell by the exact same factor, resulting in the two terms canceling out when solving for the OC condition, but remaining when calculating $I_{sc}$.

The equalities within Eq. (9) demonstrate the advantage of DC over CM, given that the currents are essentially equal to each other.

### 2.4. *Generalization to Multiple Splitting/Generation*

The equations above can be generalized to include multiplicity, *M*, of either DC layers, or CM occurrences. While the models of the previous two sections were for only a single additional splitting occurrence (*M*=2), we can generalize to *M* occurrences for both $I_{sc}$ and $V_{oc}$. Including a concentration component, *C*, we can rewrite Eq. (2) to include *M*, which is generally written as:

$$\dot{N}_{ph,in}{}^{DC/CM}(M) = gC\Omega_S \left( \sum_{m=1}^{M} m \int_{m \times E_g}^{(M-1) \times E_g} \frac{E^2 dE}{e^{E/kT_S} - 1} + M \times \int_{M \times E_g}^{\infty} \frac{E^2 dE}{e^{E/kT_S} - 1} \right) \quad (11)$$

While this is the form generally written in other works [8,17,18,23,24], we can simplify it by collecting the integrals and rewriting it as:

$$\dot{N}_{ph,in}{}^{DC/CM}(M) = gC\Omega_S \sum_{m=1}^{M} \int_{m \times E_g}^{\infty} \frac{E^2 dE}{e^{E/kT_S} - 1} \quad (12)$$

We can further rewrite this equation using a generalization of Eq. (6), for a given multiplicity $m \leq M$:

$$\theta_m \equiv \int_{m \times E_g}^{\infty} \frac{E^2 dE}{e^{E/kT_S} - 1} \bigg/ \int_{E_g}^{\infty} \frac{E^2 dE}{e^{E/kT_S} - 1} \qquad (13)$$

Note that for $m=1$, $\theta_1=1$, and for $m=2$, $\theta_2$ is identical to Eq. (6). Using Eq. (13) in Eq. (12), we obtain:

$$\dot{N}_{ph,in}^{DC/CM}(M) = gC\Omega_S \times \sum_{m=1}^{M} f_m \times \theta_m \times \int_{E_g}^{\infty} \frac{E^2 dE}{e^{E/kT_S} - 1} \qquad (14)$$

This rightmost compact form includes the original blackbody emission from the sun, multiplied by a factor $\Sigma f_m \times \theta_m \geq 1$. The efficiency for DC/CM per splitting level is given by $f_m$, with $f_1 \equiv 1$ (no splitting) and otherwise $0 \leq f_{m>1} \leq 1$. Maximal efficiency calculations are done by applying: $f_m=1$. Using the same form, we can similarly rewrite Eq. (9) as:

$$\dot{N}_{ph,out}^{CM}(new) = g \times 4\pi \times \sum_{m=1}^{M} f_m \times \theta_m \times \int_{E_g}^{\infty} \frac{E^2 dE}{e^{(E-\mu)/kT_o} - 1} \qquad (15)$$

Since the output of the DC system remains the same, following Eq. (3), the open-circuit can be found as:

$$qV_{oc}^{DC}(M) = qV_{oc}^{reg}(M) + kT_o \ln\left(\sum_{m=1}^{M} f_m \times \theta_m\right) \qquad (16)$$

Note that the entropy *gain* for the DC system is by a factor of $\Sigma\theta_m \geq 1$. In contrast the equality: $V_{oc}^{CM}=V_{oc}^{reg}$ [Eq. (10)] remains the same even when including multiplicity.

### 3. Numerical Evaluation of DC vs. CM Efficiencies

To analyze the efficiency of DC and CM systems with high multiplicity (*M*) or high concentration (*C*), the Fill-Factor approximation [35] is no longer valid, since the filling fraction ratio approaches unity at the UE limit. Therefore, the efficiency must be evaluated using a direct numerical evaluation of the power, $P=I \times V$, finding the maximum as a function of *V*, and then evaluating the efficiency at that value ($V_{max}$). Since the efficiency of a solar cell is a ratio of the electrical power to the incoming solar power, many of the constants within the formulas cancel out, leaving a single multiplication by $G=15/\pi(kT_s)^4$ [units of eV, with $kT_s=0.517$ eV]. As such, the efficiency of a DC system can be found by:

$$\eta_{eff}^{DC} = G \times V_{max} \times \left[ \int_{E_g}^{\infty} \frac{E^2 dE}{e^{E/kT_S} - 1} \times \sum_{m=1}^{M} \theta_m - \frac{4\pi}{C\Omega_S} \int_{E_g}^{\infty} \frac{E^2 dE}{e^{(E-qV_{max})/kT_o} - 1} \right] \qquad (17)$$

Whereas the CM system includes the $\Sigma\theta_m$ term outside the brackets:

$$\eta_{eff}{}^{CM} = G \times V_{max} \times \sum_{m=1}^{M} \theta_m \times \left[ \int_{E_g}^{\infty} \frac{E^2 dE}{e^{E/kT_S} - 1} - \frac{4\pi}{C\Omega_S} \int_{E_g}^{\infty} \frac{E^2 dE}{e^{(E-qV_{max})/kT_o} - 1} \right] \quad (18)$$

Once again, the CM calculation can be seen to be similar to the regular solar cell, multiplied by a factor of $\Sigma\theta_m \geq 1$. For all the calculations cited, the conversion efficiency for DC or CM was taken as unity, such that $f_{DC}=f_{CM}=1$ (or $f_m=1$). These results are therefore the maximal obtainable using either of these methods, without including any quantum efficiency losses such as we have previously done for the DC system [13]. No approximations were used when solving these equations numerically [36].

For a multiple splitting levels ($M\rightarrow 15$), the efficiency calculated using Eqs. (14) and (15), as well as that for a regular solar cell (the solution at $m=1$) have been plotted in Fig. 1. Most of the efficiency gain can be seen to occur after the first splitting event ($m=2$), with a $\approx 7\%$ efficiency gain for the DC/CM systems over a regular solar cell at optimal band-gap. While the difference between the DC and CM systems is small, this does not take into account any other losses, as will be discussed in section 5. The maximal efficiencies are tabulated in Table 1.

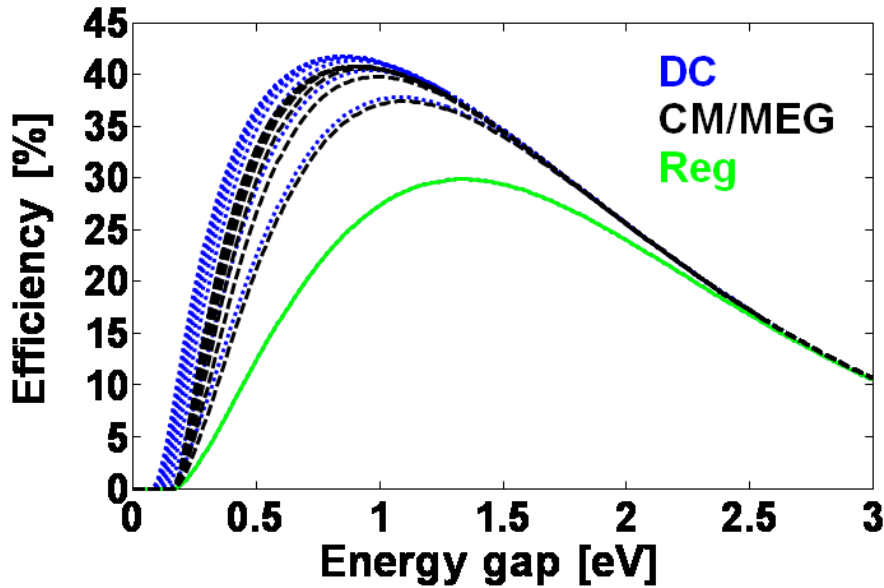

**Figure 1.** Multiple splitting levels at no concentration ($C=1$) for a regular solar cell (green solid curve); CM cell (dashed black curve); and DC cell (dotted blue curve). The efficiency increase is approximately 7% for the DC and CM cells at the first splitting level ($m=2$) with a 0.4% difference at peak efficiency. Efficiencies are plotted for $M\rightarrow 15$, with both sets of efficiencies saturating after m$\approx$3. The saturated values are higher for DC than for CM, with a peak efficiency difference of 1%.

**Table 1.** Optimal efficiency and band-gap for multiplicity with no concentration ($C=1$):

|  | Regular | | CM | | DC | |
| --- | --- | --- | --- | --- | --- | --- |
|  | $\eta_{max}$ [%] | $E_g^{max}$ [eV] | $\eta_{max}$ [%] | $E_g^{max}$ [eV] | $\eta_{max}$ [%] | $E_g^{max}$ [eV] |
| $m=1$ | 29.83 | 1.34 | - | - | - | - |
| $m=2$ | - | - | 37.4 | 1.1 | 37.76 | 1.09 |
| $m=3$ | - | - | 39.78 | 0.99 | 40.41 | 0.96 |
| $m=15$ | - | - | 40.81 | 0.91 | 41.75 | 0.85 |

Tracking the peak efficiency per multiplicity level (*M*) is plotted in Figure 2, for varying levels of concentration, ranging from *C*=1 (no concentration) to a maximal 4π concentration (for a spherical solar cell). Each point in the curve corresponds to a slightly different peak band-gap. Each set of curves (DC in blue circles and CM in black squares) can be fitted to good accuracy using a basic exponential function: $\eta(m) = \eta_{(m\to\infty)}[1-exp(-m/X)] + \eta_{(m=0)}$, and the parameter *X* is found to be the average of the peak energy-gap for *m*=1 to *m*→∞ divided by $kT_S$ (0.517 eV). This result cannot be found analytically, but it can be approximated assuming that current sums of Eqs. (14) and (15) can be approximated by a geometric series [37]. For reasonable values of concentration (*C*<1000), the majority of efficiency increase can be obtained with 1-2 splitting levels.

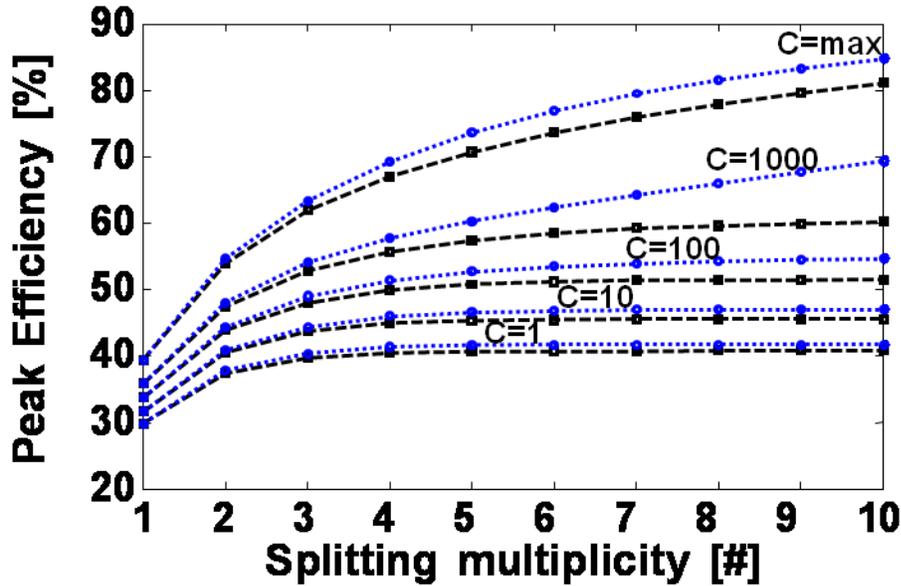

**Figure 2.** Peak efficiency for different multiplicity levels and concentrations. Each curve saturates after a few multiplicity levels.

## 4. Heat Generation Comparison

An alternative approach to comparing the methods of solar cell efficiencies is via the production of heat by the thermalization of electrons after the solar cell absorbs photons with energy above the band-gap. This thermalization loss is intentionally not included in the Ruppel-Würfel derivation [19], although modifications to the formula for $V_{oc}$ have been made to include non-radiative losses within the solar cell [25,27].

For a regular solar cell (see Fig. 3(a)), the heat loss per photon can be derived using a few basic equations. Assuming an incoming photon with energy $E=h\nu=h(\nu_g+\nu_{ex})$, with $h\nu_{ex}$ being the "excess" energy beyond the band-gap, and emitted at the band gap via band-to-band e-h radiative recombination, the difference in energy will be $\Delta E=-h\nu_{ex}$. Considering the system depicted in Fig. 3(a), the fundamental thermodynamic relation including energy, entropy and particle generation with chemical potential is:

$$dU = TdS + \mu dN \qquad (19)$$

In this system, the boundaries include the incoming and outgoing photons as particles, as well as the solar cell itself, as in Fig. 3(a). In our case, while the re-emitted photons can be considered to have a chemical potential related to the interaction with the solar cell material [38,39] (at open-circuit, this is $\mu_{oc}$), there is no net change in photons in the system, such that $dN=0$. The re-emitted blackbody photons can therefore be considered "cooler" than the solar ones [16]. Therefore, the change in entropy can be found by dividing by the temperature of the ambient (~300 K), providing the entropy/heat generated externally: $\Delta S_{lost}=h\nu_{ex}/T$.

Comparing the DC and CM cases can then be done graphically for a single splitting level as in Figs. 3(b) and (c). The DC system is once again split into two, such that the incoming photons with energy higher than $2E_g$ (purple) create thermalization heat losses in the DC layer *only*, whereas all photons between $E_g$ and $2E_g$ (blue) will create thermalization losses in the solar cell itself. Band-to-band emitted photons, including the DC photons (for a DC system with a midlevel trap site) produce no excess heat and are thus completely reversible phenomena [16]. The heat losses for all photons with $E>2E_g$ are thus limited to the (multiple) DC layers, minimizing the heating of the actual solar cell. The thermalization losses affecting $V_{oc}$ are thus limited to those occurring within the solar cell with $E_g \leq E \leq 2E_g$.

In contrast, the CM system does not have the benefit of isolating the splitting layer that will absorb the excess heat. Instead, as depicted in Fig. 3(c), all thermalization losses occur within the cell itself, and therefore add up to become the exact same of heat loss that would have occurred in a regular solar cell without any increase in current. Therefore, the CM system can be viewed as having the gain of superior current to a regular solar cell, while suffering from nearly the same heat loss generation as a regular solar cell due to thermalization. Thus, for every photon above $2E_g$ in a CM process, the thermalization loss is *half* what it would have been in a regular solar cell [33], since the loss is averaged over two e-h pairs.

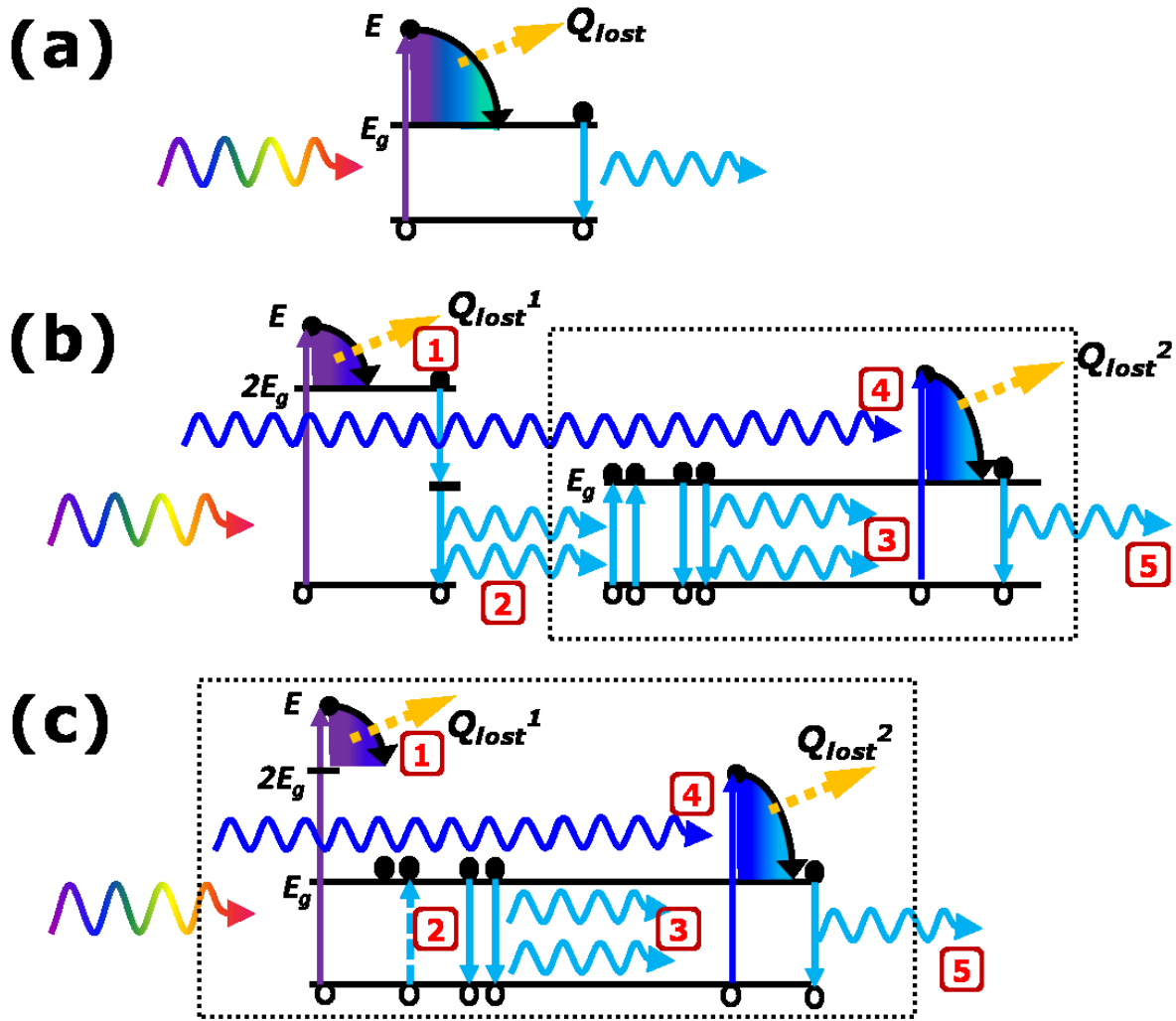

**Figure 3.** Heat transfer descriptions of regular, DC and CM solar cells. For every solar photon that goes in, a blackbody photon with chemical potential emerges (cyan) that produces no excess heat due to thermalization ("cool" photons). (a) A regular solar cell: higher energy photons (cyan-violet) produce thermalization heat ($Q_{lost}$) for all photons with energy above the bandgap. (b) DC system: photons above $2E_g$ (violet-blue, 1) create some thermalization loss, and are converted into two bandgap matched photons (cyan, 2) which produce no heat in the underlying solar cell (dotted box, 3). Photons with energy between $E_g$ and $2E_g$ (blue, 4) produce some thermalization (blue-cyan) and a single cool photon re-emitted (cyan, 5). (c) CM system converts some higher energy photons, with an associated heat loss ($Q_{lost}^1$, 1), and generates two electrons. The second electron is generated with no associated radiation (dashed cyan arrow, 2). These electrons can recombine to create two cool photons (cyan, 3). Photons with energy between $E_g$ to $2E_g$ produce thermalization loss as well ($Q_{lost}^2$, 4), while only emitting a single photon each (5).

## 5. Discussion

To compare the systems, we developed a new simplified model for CM. However, it has been established that some of the models for CM [30] are not considered thermodynamically correct, with a violation of the 2$^{nd}$ law of thermodynamics at high concentration [21,22]. Furthermore, the results of all these models appear to "break down" at low band-gap energies, below $kT_S$=0.517 eV where approximations of the integral typically collapse [23,24], leading to the non-physical result that the optimal choice of material is a zero band-gap material, with infinite multiplicity. This result can be seen by directly evaluating the UE limit, which is a simple multiplication of the voltage, $V$, by the incoming photon flux only [Eq. (14), or using $T_o$=0 K] – in that case, the UE limit approaches 100% for $M$=500 (data not shown). This UE limit surely violates the 2$^{nd}$ law of thermodynamics, since the efficiency is *beyond* the Carnot efficiency even for a photo-thermal device [16].

To compare our models with those in the literature, we will first briefly describe them: The first is the original "WKBQ model" [30] for CM, whereby the incoming flux is the summation over $M$ of Eq. (14), but the outgoing flux is simply Eq. (3), which is a single BB photon flux. This model was claimed to violate the 2$^{nd}$ law of thermodynamics [21,22] since at high concentration and low band-gap, net negative entropy is produced. The second model is the "improved WKBQ model" [17,18], in which the outgoing photon flux has been modified as well to include a summation over $M$, but also modifies Eq. (3) such that the chemical potential for each multiplicity level is multiplied by $m \times \mu$:

$$\dot{N}_{ph,out}^{MEG}(M) = qg \times 4\pi \left( \sum_{m=1}^{M-1} \int_{m \times E_g}^{(m+1) \times E_g} \frac{E^2 dE}{e^{(E-m \times \mu)/kT_o}-1} + \int_{M \times E_g}^{\infty} \frac{E^2 dE}{e^{(E-M \times \mu)/kT_o}-1} \right) \quad (20)$$

This is the generalization of Eq. (8) including multiplicity $M$>2. While this result "corrects" the apparent 2$^{nd}$ law violation, it adds the possibility of multiple internal chemical potentials [32], which can be physically interpreted as inverse Auger recombination. In terms of Kirchhoff's law of radiation, on a per-photon basis, the first model was assumed to be false, since at open-circuit equilibrium conditions, the emission from the CM system should be higher, which is why additional output emission terms were added to the improved model. The second model resulted in a maximal efficiency at maximal concentration of ~85%, which is the "blackbody limit" [8], and was thus assumed to be thermodynamically correct.

Our work offers another two models, one for the DC system (which is described a bit differently from our previous work [13]), and another for CM systems. Our CM model was designed to remove the discrepancy within the second WKBQ model, which has multiple chemical potentials, and the fact that the *useable* chemical potential at OC is the *actual* voltage measured in a solar cell [40], so that there should be no degeneracy in this definition, as there is in hot-carrier solar cells [34]. Our models are based on Kirchhoff's law of radiation, ensuring that on a per-photon basis, there is a linear equivalence of flux, as well as a thermodynamic relation using the Gibbs free energy [Eq. (1)]. It can easily be seen that our model for the DC system coincides with the original WKBQ model, which was assumed to be thermodynamically *incorrect*. However, our model for DC is based on the *correct* formulation of Kirchhoff's law, since the DC system modifies the solar spectrum, and the analysis at OC is evaluated with a modified incoming spectrum, impinging upon a solar cell that must re-emit the light that it absorbs. To accommodate for this increase in flux, $\mu$ must increase. As a result, our DC system has a *gain* factor in the $V_{oc}$ term, as established in Eqs. (7) and (16), which amounts to a reduction of the entropy created in the *solar cell*. This is perfectly plausible thermodynamically, and results in an efficiency approaching the Carnot efficiency for $M \to \infty$ and $C_{max}$. In contrast, the CM model derived by us here does not include any entropic gain, as argued by us in sections 2.2-2.4, resulting in slightly lower efficiencies that would still appear to "violate" the 2nd law of thermodynamics.

We can reconcile this problem by reinterpreting the gain factor in the entropy. We had previously claimed that the gain term can be described as a reduction in disorder of the incoming photon flux, for energies above $2E_g$ [13], however, by reevaluating equations (17) and (18), we can reassign the meaning of the $\Sigma\theta_m$ term by attaching it to the concentration term, $C$; in the formula for $V_{oc}$, this can easily done by combining the logarithmic terms as well. We can now define a *new* concentration factor as:

$$C_{new} = C_0 \times \sum_{m=1}^{M} \theta_m \tag{21}$$

with the original concentration factor $C_0$ being the one typically used, reaching 46000 suns for a flat cell geometry ($\Omega_S/\pi$ sr), or 184000 for a spherical geometry ($4\pi$ sr). This concentration term includes both the geometrical factor, which includes the solid angle of the absorption spectrum, as well as the energetic concentration that increases the current. Typically, for a flat solar cell, the concentration is limited to $2\pi$ sr [41,42] by a simple geometrical application of the 2nd law of thermodynamics. However, with multiplicity, this upper limit is surpassed due to the *increase* in illumination that effectively can be considered as a "hotter sun" (though this is not strictly correct, as the spectrum emanating from the final DC layer will be quite different, with the same *number* of photons/sec as the hotter sun would have emitted). Therefore, our model for the DC system, which includes entropic gain, is thermodynamically plausible, and can be considered a re-interpretation of the meaning of the *total* optical étendue [25] to include concentration greater than $C_0^{max}$.

It should furthermore be noted that even if using the second WKBQ model [17,18], one would still obtain a slightly lower value for $V_{oc}$ compared with that of a regular solar cell. This can be derived using Eq. (20), and would be:

$$\mu_{oc}^{DC}{}_{WKBQ} = \mu_{oc}^{DC} - kT_o \ln\left[1 + 4e^{-E_g/kT_o} \times \left(e^{\mu_{oc}/kT_o} - 1\right)\right] \tag{22}$$

However, the equation is analytically unsolvable, even for a single multiplicity layer ($m=2$).

Finally, while the difference between DC and CM systems may appear negligible, the analysis here did not include any other losses. For the DC system, one of the major losses possible in placing a DC layer above a solar cell is that part of the DC photons can be backscattered away from the solar cell [32,13], which is why previous models have claimed that the optimal location for the DC layer is *below* the solar cell [12,14,15]. While the CM system does not have this geometrical concern, the number of limitations on the material is quite large [9], with internal non-radiative recombination typically negating any benefit to be had from the purported increase in internal current. Since the DC system isolates all non-radiate losses to a material system that does not directly affect the solar cell, isolating the spectral splitting from the solar cell is beneficial when including these types of losses as well.

## 6. Conclusions

We have shown that spectral splitting in a material system that is outside the solar cell is preferable to attempting to unite this feature into the solar cell itself. The increase in efficiency, as seen directly by the difference in $V_{oc}$ as described in Eqs. (7) and (16) can equal to over a full percent for DC over CM at higher multiplicity and concentration (as shown in Fig. 2). In our analysis of these systems, we have developed a new way model for the CM system. Our model for CM is different from pre-existing models [8,17,18], but fits well into the thermodynamic method of deriving solar cell efficiencies [27] as well as complying with Kirchhoff's law of radiation. It describes a linear increase in emission due to the linear increase in absorption for CM systems, as opposed to the non-linear increase in emission in the DC system.

While the concept of uniting the spectral splitting and solar cell into a single material is appealing, the number of restrictions on CM systems recently enumerated [9,11] questions this distinction. In particular, since CM systems have been shown to be predominantly efficient in quantum dot systems, which are more difficult to contact electronically and are replete with non-radiative losses, the expected efficiency gains are not actualized. As a result, the $V_{oc}$ of these systems, which is directly related to the entropic losses, is typically lower than expected. In contrast, since a DC system is isolated from the solar cell, both systems can be optimized *independently*, with a DC system added as an additional layer to an existing solar cell. By "out-sourcing" the spectral-splitting capability to the DC layer, the entropic losses that can occur within this layer would not affect the $V_{oc}$ of the solar cell itself.

The major benefit to be had using either DC or CM would be in adding 1-3 splitting layers (or CM occurrences) to the solar cell, at moderate concentration, where heat losses at series resistance do not reduce the overall efficiency [35]. Furthermore, the advantages of spectral splitting using other means, such as up-conversion [43] can be compared to their counterparts such as interband transitions [44], which are also assumed to be somewhat similar thermodynamically [8]. We can therefore distinguish between two subsets of 3$^{rd}$ generation concepts: an "optical" approach, where the spectrum is modified externally; and a "solid-state" approach, where the spectrum is modified internally. Attempting to improve the efficiency of solar cells by modifying the spectrum externally, and isolating associated entropic losses, was here demonstrated to be thermodynamically preferable.

**Acknowledgments**

ZRA thanks Prof. Viorel Badescu for his helpful discussion. This work was supported by the U.S. Department of Energy, Basic Energy Sciences Energy Frontier Research Center (DoE-LMI-EFRC) under award DOE DE-AC02-05CH11231 and by the National Science Foundation Nano-Scale Science and Engineering Center (NSF-NSEC) under award CMMI-0751621. ZRA acknowledges the National Defense Science and Engineering Graduate (NDSEG) Fellowship, 32 CFR 168a.